\documentstyle[12pt,titlepage]{article} % Create extra title page
\title{The Isgur-Wise function in a relativistic model for $q\bar Q$ system}
\author{Mohammad R. Ahmady, Roberto R. Mendel and James D. Talman}
\date{July, 1994}   % Deleting this command produces today's date.
\def\_{\rule{.3em}{.15ex}}  % Get underscore by typing \_.

\begin{document}           % End of preamble and beginning of text.
\begin{titlepage}
%\begin{flushright}
% IC/92/404\\
% HEP-PH 9211332
%\end{flushright}
 \begin{center}
  \vspace{0.75in}
  {\bf {\LARGE The Isgur-Wise function in a relativistic model for
  $q\bar Q$ system} \\
  \vspace{0.75in}
  Mohammad R. Ahmady, Roberto R. Mendel and James D. Talman} \\
  Department of Applied Mathematics\\
  The University of Western Ontario \\
  London, Ontario, Canada\\
  \vspace{1in}
  ABSTRACT \\
  \vspace{0.5in}
  \end{center}
  \begin{quotation}
  \noindent
We use the Dirac equation with a ``(asymptotically free) Coulomb + (Lorentz
 scalar) linear '' potential to estimate the light quark wavefunction for $
 q\bar Q$ mesons in the limit $m_Q\to \infty$.  We use these wavefunctions
 to calculate the Isgur-Wise function $\xi (v.v^\prime )$ for orbital and
 radial ground states in the phenomenologically interesting range $1\leq v.v^
 \prime \leq 4$.  We find a simple expression for
 the zero-recoil slope, $\xi^\prime
 (1) =-1/2- \epsilon^2 <{r_q}^2>/3$, where $\epsilon$ is the energy
 eigenvalue of the light quark, which can be identified with the $\bar\Lambda
 $ parameter of the Heavy Quark Effective Theory.  This result implies an
 upper bound of $-1/2$ for the slope $\xi^\prime (1)$.  Also, because for a
very
light quark $q (q=u, d)$ the size $\sqrt {<{r_q}^2>}$ of the meson is
determined mainly by the ``confining'' term in the potential $(\gamma_\circ
\sigma
 r)$, the shape of $\xi_{u,d}(v.v^\prime )$ is seen to be mostly sensitive
 to the dimensionless ratio $\bar \Lambda_{u,d}^2/\sigma$.  We present results
 for the ranges of parameters $150 MeV <\bar \Lambda_{u,d} <600 MeV$
  $(\bar\Lambda_s \approx \bar\Lambda_{u,d}+100 MeV)$, $0.14 {GeV}^2 \leq
  \sigma \leq 0.25 {GeV}^2$ and light quark masses $m_u, m_d \approx 0,
  m_s=175 MeV$ and compare to existing experimental data and other theoretical
  estimates.  Fits to the data give:
  ${\bar\Lambda_{u,d}}^2/\sigma =4.8\pm 1.7
$, $-\xi^\prime_{u,d}(1)=2.4\pm 0.7$ and
 $\vert V_{cb} \vert \sqrt {\frac {\tau_B}{1.48 ps}}=0.050\pm 0.008$
 [ARGUS '93];
  ${\bar\Lambda_{u,d}}^2/\sigma =3.4\pm 1.8
$, $-\xi^\prime_{u,d}(1)=1.8\pm 0.7$ and
 $\vert V_{cb} \vert \sqrt {\frac {\tau_B}{1.48 ps}}=0.043\pm 0.008$
 [CLEO '93];
  ${\bar\Lambda_{u,d}}^2/\sigma =1.9\pm 0.7
$, $-\xi^\prime_{u,d}(1)=1.3\pm 0.3$ and
 $\vert V_{cb} \vert F(1)=0.037\pm 0.003$
 [CLEO '94] (Existing theoretical estimates for $F(1)$ fall in the range
 $0.86<F(1)<1.01$).  Our model seems to favour the CLEO '94 data set in two
 respects: the fits are better and the resulting ranges for the model
 parameters ($\bar\Lambda_{u,d}$, $\sigma$) are more in line with independent
 theoretical estimates.

\end{quotation}
\end{titlepage}

%\maketitle                 % Produces the title.
%\tableofcontents          % Must follow by a \newpage
%\newpage
%\includeonly{file1}       % Cause the argument in \include{} to be processed
%\include{file1}
%\input{your_file}

\newcommand{\da}{\mbox{$\scriptscriptstyle \dag$}}
\newcommand{\lag}{\mbox{$\cal L$}}
\newcommand{\tr}{\mbox{\rm Tr\space}}
\newcommand{\fc}{\mbox{${\widetilde F}_\pi ^2$}}
\newcommand{\ns}{\textstyle}
\newcommand{\si}{\scriptstyle}

\section{Introduction and Summary}
During recent years, a lot of effort \cite {1} has gone into the
description of systems and processes involving at least one heavy quark
$(m_Q \gg \Lambda_{QCD})$ via a systematic expansion in the small parameters
 $(\Lambda_{QCD}/m_Q)$ and $\alpha_s (m_Q^2)$, taking advantage of ``Heavy
 Quark Symmetries''.  The usefulness of this approach is in that in the heavy
 quark limit $m_Q\to \infty$, all the physics can be expressed in terms of
 a small number of form factors, which depend on the light quark and gluon
  dynamics only.  These ``universal'' functions can be used as a means for
   comparison among different theoretical models (such as non-relativistic
 and relativistic potential models, lattice QCD calculations, etc).
Comparisons
 with experiment or with methods such as QCD sum rules require in general
 introducing model dependent $O(\frac {\Lambda_{QCD}}{m_Q})$ corrections (
 except at $v.v^\prime =1$, where corrections start at order $({\Lambda_{QCD}
 }^2/m_Q^2)$ \cite{LUKE}), as well as calculable perturbative QCD corrections.

 In this paper we limit ourselves to the leading order in the heavy quark
expansion.  We calculate the Isgur-Wise function \cite {2}, $\xi (v.v^\prime
)$, for radial and orbital ground state mesons using a relativistic model
for the light quark.  This function in our case (radial and orbital ground
state), for a given flavour of light quark, fully describes the transition
$M_q\to M^\prime _q$ in which a local operator transforms the heavy quark
$Q$ in the initial meson $M_q$ (of 4-velocity $v$ ; J=0 or 1) into the heavy
quark $Q^\prime$ in the meson $M_q^\prime$ (of 4 velocity $v^\prime$ ;
J=0 or 1).

We assume that the light quark wavefunction obeys a Dirac equation with a
spherically symmetric potential in the reference frame in which the heavy
quark is stationary at the origin.   We also assume that the potential has
 the form $V(\vert \vec x\vert )=V_c (\vert \vec x \vert )+c_\circ +\gamma
^\circ \sigma \vert \vec x \vert$, where $V_c (\vert \vec x \vert ) $ is
an asymptotically free Coulomb term, $c_\circ$ is a constant and the last
term is a Lorentz scalar confining term.  Thus, the spatial wavefunction for
the light quark q, of mass $m_q$, is assumed to obey the time-independent
Dirac equation
\begin{equation}
\left [ \vec\alpha .(-i\vec\bigtriangledown_{\vec x}) +
V_c (\vert \vec x \vert ) +c_\circ +\gamma^\circ (\sigma \vert \vec x \vert
+m_q) \right ] \Psi (\vec x )=\epsilon \Psi (\vec x ) \; .
\end{equation}
Because we are investigating the meson system in the heavy quark limit
$(\Lambda_{QCD}/m_Q \to 0)$, the energy eigenvalue $\epsilon$ of the Dirac
equation can be identified with the ``inertia'' parameter $\bar\Lambda_q$
often introduced in the Heavy Quark Effective Theory [HQET] \cite {1}
\begin{equation}
\epsilon \approx \bar \Lambda_q \equiv \lim_{M_Q\to\infty}(M_{{(q\bar Q)}_
{meson}}-M_Q)
\end{equation}

In our investigation, we allow the $\bar\Lambda$ parameters to vary over the
range $0.15\leq \bar\Lambda_{u,d}\leq 0.6 GeV$ $(\bar\Lambda_s \approx \bar
\Lambda_{u,d}+100 MeV)$, obtained from recent lattice gauge theory studies
\cite {3} and other theoretical estimates \cite {4,5}.  As will be seen below,
 the parameter $\bar\Lambda$ plays an important role in the calculation of the
Isgur-Wise function.  The additive constant $c_\circ$ in the potential
allows us to pick an arbitrary value for $\bar\Lambda$.  The asymptotically
free Coulomb term $V_c(r)$ is parametrized in terms of the QCD scale $\Lambda
_{\overline {MS}}$ and a saturation value for the strong coupling
$\alpha^\infty
_s \equiv \alpha_s(r=\infty )$ (We use $\Lambda_{\overline {MS}}=240 MeV$ and
$\alpha_s^\infty\approx 1$).  Phenomenological and theoretical considerations
 motivate our use of 3 different values for the string tension : $\sigma =
 0.25 {GeV}^2, 0.18 {GeV}^2$ and $0.14 {GeV}^2$.  We set $m_u, m_d =0$ and
 $m_s=0.175 GeV$.  Further details about the parametrization of the potential
  are given later in the paper.

We find that the shape of the Isgur-Wise function is mostly sensitive to
the parameters $\bar\Lambda$ and $\sigma$.  In fact, to a very good
approximation
 $\xi_{u,d}(v.v^\prime )$ depends on the dimensionless ratio $\frac {{\bar
 \Lambda}^2}{\sigma}$ only.  Within our approximation, we also find a strict
  upper bound on the slope at zero recoil: $\xi^\prime (1)<-1/2$.  A recent
   paper \cite {6} that uses the MIT Bag Model formalism obtains the same
 bound, which is stronger than the well known Bjorken bound
 $\xi^\prime (1)<-1/4$ \cite{7}.  For large values of $v.v^\prime$
 ($2<v.v^\prime
 <4$) the shape of $\xi (v.v^\prime )$ is relatively insensitive to the
 input parameters $\bar\Lambda$, $\sigma$ and $\Lambda_{\overline {MS}}$.

 We present explicit results for $\xi (v.v^\prime )$ in the
 phenomenologically interesting region $1\leq v.v^\prime \leq 4$, using the
 above mentioned range of parameters.  We compare our results with recent
 experimental data \cite {8,9,CLEO94} of semileptonic B decays as well
 as with other theoretical estimates.  From fits to the ARGUS '93 data
 \cite {8} we obtain the ranges
 $\frac {{\bar\Lambda_{u,d}}^2}{\sigma} =4.8\pm 1.7$ (corresponding to
 $-\xi^\prime_{u,d} (1) =2.4\pm 0.7$) and
 $\vert V_{cb}\vert \sqrt {\frac {\tau_B}{1.48 ps}}=0.050\pm 0.008$.  Although
  the fits are of good quality, for reasonable values of the string tension
  $\sigma$ they favour values for the inertia parameter $\bar\Lambda_{u,d}$
  that are significantly above the range $0.15 GeV<\bar\Lambda_{u,d}<0.6 GeV$
advocated by most theoretical estimates \cite {3,4,5}.  The corresponding
 range for the slope $-\xi_{u,d} (1)$ is also above most independent
theoretical estimates (see discussion in section 4).  The best fit to the
CLEO '93 data \cite {9} was of significantly poorer quality.  Here
we found the ranges
 $\frac {{\bar\Lambda_{u,d}}^2}{\sigma} =3.4\pm 1.8$ (corresponding to
 $-\xi^\prime_{u,d} (1) =1.8\pm 0.7$) and
 $\vert V_{cb}\vert \sqrt {\frac {\tau_B}{1.48 ps}}=0.043\pm 0.008$.
 The ranges for $\frac {{\bar\Lambda_{u,d}}^2}{\sigma}$ has in this case some
overlap with previous theoretical estimates, but is still somewhat on the
high side of these estimates.  We noticed that if the data point from the
CLEO '93 set \cite {9} corresponding to the largest value of $v.v'$ is
ignored, the quality of the fit greatly improves.  The parameter ranges
obtained in this case are
 $\frac {{\bar\Lambda_{u,d}}^2}{\sigma} =2.1\pm 1.6$ (corresponding to
 $-\xi^\prime_{u,d} (1) =1.3\pm 0.6$) and
 $\vert V_{cb}\vert \sqrt {\frac {\tau_B}{1.48 ps}}=0.038\pm 0.006$.  These
 ranges for $\frac {{\bar\Lambda_{u,d}}^2}{\sigma}$ and $-\xi'_{u,d}
(1)$ overlap with many
previous theoretical estimates, but are somewhat too wide to provide useful
new information that could distinguish between these.

The recently released CLEO '94 data analysis \cite {CLEO94} has smaller
error bars than the data mentioned above \cite {8,9}.  Also, our model
produces better quality fits to this data set than to the previous ones.  These
two factors contribute to help narrow the parameter ranges.  The following
ranges are favoured:
 $\frac {{\bar\Lambda_{u,d}}^2}{\sigma} =1.9\pm 0.7$ (corresponding to
 $-\xi^\prime_{u,d} (1) =1.3\pm 0.3$) and
 $\vert V_{cb}\vert F(1)=0.037\pm 0.003$ (Different theoretical estimates for
 the constant $F(1)$ can be found in eq (40), section 4).  Note that the
 central values of these ranges are quite close to the ones found from the
 fits to the CLEO '93 data set after removing the CLEO '93 data point
 corresponding to the highest $v.v'$.  The uncertainties, however, are here
 reduced by a factor of about 2, which should help us to narrow down the
 parameter space of the underlying physics.

 \section{General Formalism}
 We assume that the wavefunction $\Psi$ of a $q\bar Q$ meson of mass $
 M_{q\bar Q}$ in the limit $m_Q\to\infty$ (i.e ignoring $O(\frac {\Lambda_
 {QCD}}{m_Q})$ effects) can be expressed in terms of a direct product of the
(free spinor) wavefunction of the heavy anti-quark, $\chi$, and the
wavefunction of the light quark, $\psi$, in the ``relative''
coordinates.  It is therefore convenient to define the 4-vectors
\begin{equation}
X^\mu \equiv (t_Q,\vec r_Q) \; , \; x^\mu \equiv (t_q-t_Q , \vec r_q
-\vec r_Q)
\end{equation}
Of course, the kinematics of the two wavefunctions $\chi$ and $\psi$
are not completely independent.  They are connected by the implicit
constraint that in the rest frame of the heavy quark (which is also
the rest frame of the $q\bar Q$ meson) the spatial part of the light
quark wavefunction, $\psi (\vec x )$, obeys the time-independent
Dirac equation (1).  In order to be able to give the meson
wavefunction $\Psi$ a physical meaning in a given reference frame, we
have to set the ``relative time'' of the two constituents to zero
%\begin{equation}
$$x^\circ =0 \; , \; t_q=t_Q \equiv t$$
%\end{equation}

In the rest frame of the meson ($\vec P=0$), its wavefunction can
then be written as:
\begin{equation}
\Psi_{\vec 0}^{\lambda , \eta }(X,x) = \sqrt {\frac {2M_{q\bar
Q}}{{(2\pi )}^3}} \left [\chi_{\vec 0}^\lambda e^{-im_Qt_Q}\right ]\otimes
\left [ \psi_{\vec 0}^\eta (\vec x ) e^{-i\epsilon t_q}\right ]\; ,
\end{equation}
where $\lambda\; ,\; \eta$ are spin indices ($\lambda \; ,\; \eta = \uparrow
{\rm or}\downarrow$ with respect to some axis); $\chi^\lambda_{\vec 0}$ is
a free Dirac spinor at rest corresponding to the heavy antiquark and $\psi^
\eta_{\vec 0} (\vec x)$ obeys equation (1) with energy eigenvalue $\epsilon$.
  The factor in front of the direct product is for normalization purposes
  (See eq (10) below).  Using our convention $t_Q=t_q=t$ we see that the rest
mass $M_{q\bar Q}$ is given by
\begin{equation}
M_{q\bar Q} =m_Q +\epsilon
\end{equation}
thus validating our identification of $\epsilon$ with the parameter $\bar
\Lambda_q$ commonly introduced in HQET.

We now turn to the description of the meson $q\bar Q$ as seen from a
reference frame $L$, with respect to which the meson (rest frame
$L^\prime$) is moving with a 4-velocity $v\equiv (\gamma , \gamma
\vec \beta )$.  In the unprimed frame, the 4-momentum of the meson is
given by $P=vM$.  We assume that the wavefunctions of the light quark and
heavy antiquark transform in the standard way under a Lorentz transformation.
  Although this would be strictly correct only if the Dirac equation for the
light quark (1) were in fact covariant under Lorentz transformations, we
assume (and verify in some special cases later) that this procedure will not
introduce important errors in our results.  We can then express our
wavefunctions in the unprimed frame in terms of those in the rest frame
of the meson, $L^\prime$
\begin{equation}
\begin{array}{l}
\psi_v(t_q,\vec r_q)=S_v\psi_{\vec 0}(t^\prime_q ,{\vec r_q}^\prime )\\
\chi_v (t_Q,\vec r_Q)=S_v \chi_{\vec 0}(t^\prime _Q,{\vec r_Q}^\prime )
\end{array}
\end{equation}
where
\begin{equation}
\begin{array}{l}
t_{q,Q}^\prime =\gamma \left [ t_{q,Q} -\vec \beta . \vec r_{q,Q}\right
]\\
\vec r_{q,Q}^\prime = \vec r_{q,Q}+(\gamma -1)(\hat \beta . \vec
r_{q,Q})\hat\beta -\gamma \vec \beta t_{q,Q}\\
t_q=t_Q\equiv t
\end{array}
\end{equation}
and
\begin{equation}
S_v=exp \left \{ \frac {{tanh}^{-1} \vert \vec \beta\vert}{2}\vec\alpha
.\hat\beta\right\} =\sqrt {\frac {\gamma +1}{2}}
\left [ \begin{array}{llll}
1 & 0 & \frac {\gamma\beta_z}{\gamma +1} & \frac {\gamma (\beta_x-i\beta_y)}
{\gamma +1} \\
0 & 1 & \frac {\gamma (\beta_x +i\beta_y)}{\gamma +1} & -\frac {\gamma \beta
_z}{\gamma +1} \\
\frac {\gamma \beta_z}{\gamma + 1} & \frac {\gamma (\beta_x -i\beta_y) }
{\gamma +1} & 1 & 0 \\
\frac {\gamma (\beta_x +i\beta_y )}{\gamma +1} & -\frac {\gamma \beta_z}
{\gamma +1 }& 0 & 1
\end{array} \right ] \; .
\end{equation}
Therefore, the wavefunction of a meson moving with velocity $\vec \beta$ in
our (unprimed) frame will be
\begin{equation}
\begin {array}{l}
\Psi_{\vec v}^{\lambda , \eta }(X,x) = \sqrt {\frac {2M_{q\bar
Q}}{{(2\pi )}^3}} \left [S_v\chi_{\vec 0}^\lambda e^{-im_Qt_Q^\prime}
\right ]\otimes
\left [ S_v\psi_{\vec 0}^\eta ({\vec x}^\prime ) e^{-i\epsilon t_q^\prime
}\right ]\\
= \sqrt {\frac {2M_{q\bar Q}}{{(2\pi )}^3}} e^{-iP.X} \left [S_v\chi^\lambda _
{\vec 0}\right ] \otimes \left [ S_v \psi^\eta_{\vec 0}(\vec x +(\gamma -1)
(\hat\beta .\vec x)\hat\beta )e^{i\epsilon\gamma\vec\beta .\vec x}\right ]
\; ,
\end{array}
\end{equation}
where we have used eq (7) for the last step and $t_q=t_Q=t$, $\vec r_Q=
\vec X \; , \; \vec r_q -\vec r_Q =\vec x \; , \; P=(\gamma M, \gamma
\vec \beta M) $.

It is straightforward to check that the standard normalization for meson
states,
 \begin{equation}
<\Psi_{v^\prime}^{\lambda , \eta^\prime}\vert\Psi_v^{\lambda , \eta}>=
2P^\circ\delta^3(\vec P -{\vec P}^\prime )\delta_{\lambda \lambda^\prime}
\delta_{\eta \eta^\prime} \; ,
\end{equation}
is obtained provided that the usual normalization is
used for the light and heavy quark wavefunctions:
\begin{equation}
{\chi^\lambda_{\vec 0}}^\dagger \chi^{\lambda^\prime}_{\vec 0}
=\delta_{\lambda
\lambda^\prime} \; {\rm and} \; \int d^3 \vec x {\psi^\eta_{\vec 0}}^\dagger
(\vec x )\psi^{\eta^\prime}_{\vec 0} (\vec x) =\delta_{\eta\eta^\prime} \; .
\end{equation}

The Isgur-Wise function $\xi (v.v^\prime )$ (for orbital and radial ground
states) can be extracted in a simple way \cite {1,2} by calculating the
following matrix element between two pseudoscalar states of mass $M$ and
4-velocities $v$ and $v^\prime$:
\begin{equation}
\xi (v.v^\prime )=\frac {1}{(1+v.v^\prime )M}<P_{v^\prime} \vert \bar h_{
v^\prime}(0)h_v(0)\vert P_v> \; ,
\end{equation}
where $\vert P_v>=\frac {1}{\sqrt 2}\left (\vert \Psi_v^{\uparrow\downarrow}
>-\vert\Psi_v^{\downarrow \uparrow}>\right )$ can be obtained from eq (9)
and $\bar h_{v^\prime} (0) (h_v (0))$ is the creation (annihilation) operator
for
a heavy quark of 4-velocity $v^\prime (v)$ at the origin of space-time and
$v^\mu\gamma_\mu h_v=h_v$ has been used.  The light quark wavefunctions $\psi
_{\vec 0}^\eta $ (see eq (9)) correspond in this case to the radial and
orbital ground state solutions of the Dirac equation (1).

Without loss of generality we can assume that $\vec\beta$ and ${\vec\beta}^
\prime$ are collinear.  We can then use the simple result
\begin{equation}
{\chi_{\vec 0}}^\dagger {S_{v^\prime}}^\dagger \gamma^\circ
S_v\chi_{\vec 0}=\sqrt {\frac
{1+v.v^\prime}{2}}
\end{equation}
to obtain the explicit expression
\begin{equation}
\xi (v.v^\prime )=\sqrt {\frac {2}{1+v.v^\prime }}\int d^3 \vec x \frac {1}
{2} \sum_{\eta =\uparrow , \downarrow}{\psi^\eta_{\vec 0}}^\dagger ({\vec x}^
\prime ){S_{v^\prime}}^\dagger S_v\psi^\eta_{\vec 0} ( {\vec x}'')e^{i\epsilon
(\gamma\vec\beta -\gamma^\prime {\vec\beta}^\prime ).\vec x}\; ,
\end{equation}
where
\begin{equation}
\begin{array}{l}
{\vec x}^\prime =\vec x +(\gamma^\prime -1)({\hat \beta}^\prime .\vec x ){
\hat\beta}^\prime \\
 {\vec x}'' =\vec x +(\gamma -1) (\hat\beta .\vec x)\hat\beta
\end{array}
\end{equation}
and $v.v^\prime =\gamma\gamma^\prime [ 1-\vec\beta .{\vec\beta}^\prime ]$
($v.v^\prime \; {\rm in} \; [ 1, \infty ]$).  ${\vec x}'$ and $ {\vec
x}''$ represent
the spatial coordinates in the reference frames where the mesons of respective
4-velocities $v^\prime$ and $v$ are at rest.

In principle, the above integral expression (14) should depend on $v.v^\prime
$ only, i.e. should be Lorentz invariant so that its value be independent
from the (unprimed) frame that is chosen to evaluate the integral in.  In
practice, because the Dirac eq (1) is not Lorentz covariant (due to the
potential), the wavefunctions are not either.  Therefore, the function $
\xi (v.v^\prime )$ does depend on the frame we choose.  However, it can be
easily checked that the important properties $\xi (1)=1$ and $Im (\xi (v.
v^\prime ))=0$ are satisfied in any Lorentz frame.   Moreover, as will be
detailed at the end of this section, we found that the value of the slope
at zero recoil $\xi^\prime (1)$ is the same in three simple (but quite
different) Lorentz frames.  Thus, at least in the vicinity of the zero recoil
point, the effects of Lorentz non-invariance of our function $\xi$ are
expected to be small.

We choose to work in the Breit \cite {6} frame, where it is easiest to
perform the calculation.  In this frame the incoming and outgoing mesons are
moving with equal speeds but in opposite directions ($\vec\beta =-{\vec\beta}^
\prime$) and therefore $v.v^\prime =2\gamma^2 -1$.  Also, $S^\dagger
_{v^\prime}
=S^{-1}_v$ so that after a change in integration variable $\xi (v.v^\prime )$
 can be written in the simple form \cite {10}, with $\vec \beta$ and
${\vec\beta
 }'$ in the z direction,
\begin{equation}
\xi_q (v.v^\prime )=\frac {1}{\gamma^2}\int d^3 {\vec r}^\prime \frac {1}{2}
\sum_{\eta =\uparrow \downarrow }{\left \vert \psi^\eta_{\vec 0} ({\vec r}^
\prime )\right\vert }^2 e^{2i\epsilon_q\beta z^\prime}\; ,
\end{equation}
where
\begin{equation}
\gamma =\sqrt {\frac {1+v.v^\prime }{2}}\; ,\; \beta =\sqrt {\frac {v.v^\prime
 -1}{v.v^\prime +1}}\; .
\end{equation}

Because $
\sum_{\eta =\uparrow \downarrow }{\left \vert \psi^\eta_{\vec 0} ({\vec x}^
\prime )\right\vert }^2$ is spherically symmetric, the angular integration
is trivial.  Also, it is easy to derive an expression for the slope at zero
recoil $\xi^\prime (1)$ in terms of the inertia parameter $\epsilon_q =
\bar \Lambda_q$ and ${<r_q^2>}_{\vec 0}$:
\begin{equation}
\frac {\partial \xi_q}{\partial (v.v^\prime )} |_{v.v^\prime =1} =-\frac {1}
{2} -\frac {1}{3} \epsilon^2_q {<r_q^2>}_{\vec 0}=
-\frac {1}{2} -\frac {1}{3} {\bar \Lambda}^2_q {<r_q^2>}_{\vec 0}\; .
\end{equation}

This is a general result within our formalism and not linked to any
particular form of the potential used in the Dirac equation for the light
quark.  It tells us that $\xi^\prime (1)$ depends only on the light quark
energy eigenvalue $\epsilon$ (or equivalently for our model, the HQET
``inertia'' parameter $\bar \Lambda_q=\lim_{m_Q\to\infty}(M_{Q\bar q}-m_Q)$)
 and on the rms distance between the light quark and the (stationary) heavy
 antiquark, $\sqrt{{<r_q^2>}_{\vec 0}}$.  We should point out that this same
result was obtained independently in a recent paper that describes $q\bar Q$
 mesons and $qqQ$ baryons in the context of the MIT Bag Model \cite {6},
  also obtained in the Breit frame.

  An interesting aspect of this result is that it sets an upper bound on the
  slope ($\xi^\prime (1)=-\rho^2$)
  \begin{equation}
 \xi^\prime (1) <-\frac {1}{2}
 \end{equation}
 This bound is stronger than the well known Bjorken bound $\xi^\prime (1) <-1/4
 $ and is a result of the dynamical assumptions inherent in our treatment
 of the meson system in the heavy quark limit.  In particular, we think that
 it can be traced to the fact that we describe a moving meson by boosting
 the light and heavy quark independently (by the same amount).

 Because our formalism is not fully covariant (spherically symmetric Dirac
 equation potential is put in by hand) one may suspect that the above result
 for $\xi^\prime (1)$ (eq 18) is Lorentz frame dependent.  We have checked
 explicitly that the result is unchanged if $\xi^\prime (1)$ is calculated
 in the frames where either the incoming or the outgoing meson is at rest.

\section{Parametrization of the Dirac equation potential}

In the $M_Q\to\infty$ limit, the $q\bar Q$ meson system should be well
described by the heavy antiquark stationary at the origin and the light quark
(of mass $m_q$) moving in a spherically symmetric static external potential.
 We are of course ignoring the self interactions of the light quark in the
hope that these can be absorbed to some extent in a renormalization of the
parameters of the external potential.

We take the short distance behaviour of the potential from
renormalization-group-improved QCD perturbation theory and the long distance
behaviour from lattice and other non-perturbative studies that ignore
screening by light quark pair creation.  Unfortunately, knowledge about the
leading behaviour of the potential in these two extreme distance regimes
defines the potential only up to an additive constant.

At short distances, the usual asymptotically free Coulomb form (transforming
 as the zeroth component of a Lorentz 4-vector) is obtained:
 \begin{equation}
V_c(r)=-\frac {4}{3}\frac {\alpha_s(r)}{r}
\end{equation}
where $\alpha_s(r)$ is obtained in the leading log approximation and is
parametrized as follows
\begin{equation}
\alpha_s(r)=\frac {2\pi}{(11-\frac {2N_F}{3})\ln [ A+\frac {B}{r}]}
\end{equation}
The parameter A defines the ``long distance'' saturation value for $\alpha
_s$.  We use $A=2$ which corresponds to $\alpha_s(r=\infty )=1.0$.  The
parameter B is related to $\Lambda_{\overline {MS}}$ by
$B={(2.23\Lambda_{\overline
{MS}})}^{-1}$ for $N_F=3$.  We use $N_F=3 \cite{11}$ throughout
because for most
distance regimes relevant to our calculation the $c\bar c$ and $b\bar b$
vacuum polarization contribution should be negligible.  We generally use the
present experimental average $\Lambda_{\overline {MS}}\approx 0.240$ GeV
\cite{12}
, which
corresponds to $B=1.87 {GeV}^{-1}$.  As described in the results section,
we do vary the value  of B (i.e. of $\Lambda_{\overline {MS}}$) with respect
to the above value for the specific purpose of studying the sensitivity
of the shape of $\xi (v.v^\prime )$ to this parameter.  We find only a weak
dependence.

To describe the long distance behaviour we use a linear term in the potential
that transforms as a Lorentz scalar (mass-like).  Many theoretical and
phenomenological arguments seem to favour this form \cite {13,14,15p},
\begin{equation}
V_L=\gamma^\circ \sigma r \; ,
\end{equation}
where $\gamma^\circ$ is the usual Dirac matrix, rather than an admixture
with a linearly rising $0^{\rm th}$ component of a vector term.  We do our
calculations with three different values for the string tension parameter
$\sigma$.  The choices that we made were arrived at as follows.  The
experimental information available for the $D$ and $D_s$ systems seems
to indicate that the (spin averaged) splitting between P states and S states
is in both cases approximately 0.45 GeV.  We found that in order to obtain a
splitting of this magnitude with our $V_c(r)+c_\circ +\gamma^\circ \sigma
 r$ as defined above, $\sigma$ had to be about $\sigma\approx 0.25 {GeV}^2$
 .  A previous study \cite {13} agrees with this calculation.  On the other
 hand, this value for $\sigma$ is significantly larger than those obtained
 from the Regge slope data ($q\bar q$ systems) \cite {14,15p}.  Of course,
  one cannot exclude the possibility that the 2P-1S splitting of about
  0.45 GeV holds for the $D$, $D_s$ systems but is significantly smaller
in the hypothetical limit $M_Q\to\infty$, which are the systems we are trying
to describe in the present work.  The difference would be due to heavy quark
recoil effects in the charmed systems.  In fact, preliminary evidence was
reported recently for a $B^{**}(\ell =1)$ candidate with mass $M_{B^{**}}
\approx 5610 MeV$ (Ref. \cite {15} and also talk by V. L\"{u}th at the same
conference), which would indicate a 2P-1S splitting of only about 0.34 GeV.
  If confirmed, this would allow us to use more conventional values for
  $\sigma$, as extracted from the Regge slope $\alpha^\prime$.  We therefore
also consider this possibility.  The extraction of $\sigma$ from $\alpha^
\prime$ is somewhat model dependent.  Two common relations are
\begin{equation}
\sigma = \frac {1}{2\pi\alpha^\prime} \;\;\;\;\; {\rm (string \;\; model
\; \cite
{16})}
\end{equation}
and
\begin{equation}
%\begin{array}{l}
\sigma =\frac {1}{8\alpha^\prime}\;\;\;\;\;
{\rm  (2-body \; generalization
\; of \; Klein \; Gordon \; Eq. \; \cite {14,15p})}
%\end{array}
\end{equation}
Using $\alpha^\prime\approx 0.9 {GeV}^{-2}$ \cite {16} we obtain respectively
$\sigma \approx 0.18 {GeV}^2$ (string model) and $\sigma\approx 0.14 {GeV}
^2$ (2-body K.G. Eq.).  The latter value agrees also with a recent lattice
estimate \cite {17}.  We would like to remark that if the 2P-1S splitting for
the B mesons (and therefore also in the $m_Q\to\infty$ limit) turns out to
be similar in magnitude to the observed splitting in the $D$, $D_s$ systems,
 as predicted in several models \cite {18} we could still fit this splitting
 using the more conventional values of $\sigma$ (0.18 and 0.14 ${GeV}^2$),
  provided that we change the $\Lambda^{(3)}_{\overline {MS}}$ parameter in
$V_c(r)$ (See equation (21) and discussion below) to
$\Lambda^{(3)}_{\overline {MS}
}\geq O(0.5 GeV)$ , i.e. $B<O(0.9 {GeV}^{-1})$.

So far we have used as input the well known leading short distance and long
distance behaviour of the potential.  There is less theoretical knowledge
about the shape of the potential in the intermediate region.  We introduce
an additive constant term $c_\circ$, which is clearly subleading both in the
short and long distance regimes.  Because the only role of this constant is
to define the absolute scale of the light quark energy $\epsilon_q$, this
constant gets absorbed once we identify $\epsilon_q\equiv \bar \Lambda_q
=\lim_{m_Q\to\infty} (M_{q\bar Q}-m_Q)$ and assign $\bar \Lambda_q$ some
physical value.

We obtain a plausible range for the physical parameter $\bar \Lambda_q$
from previous theoretical works that estimated the value of the ``pole mass''
of the b quark, $m_b$.  We use the relation
\begin{equation}
M_B \; ({\rm Spin\; averaged})\approx 5310 MeV\approx \bar \Lambda_{u,d}+
m_b+\frac {<{\vec P_q}^2>}{2m_b} \; ,
\end{equation}
where $<{\vec P_Q}^2>\approx <{\vec P_q}^2>$ has been used.

A recent lattice study finds $m_b=4950\pm150 MeV$ \cite {3}, which also agrees
 with a HQET estimate \cite {4}.  QCD sum rule estimates are lower, closer to
 $m_b\approx 4.6 GeV$\cite {5}.  From these values for $m_b$ and eq (25)
 we then find
 the range
\begin{equation}
\bar \Lambda_{u,d}\approx 150-600 MeV \; .
\end{equation}

We do our calculations using mostly the two extreme values as well as the
average value in this range $\bar \Lambda_{u,d}=150,\; 375\; {\rm and}\;
600 MeV$.  Notice that we have to make a distinction between $\bar \Lambda
_{u,d}$ and $\bar \Lambda_s$, but this does not introduce any complications
because experimentally
\begin{equation}
M_{D_s}-M_D\approx M_{B_s}-M_B\approx 100 MeV \; ,
\end{equation}
\begin{equation}
M_{D^*_s}-M_{D_s}\approx M_{D^*}-M_D\; ,
\end{equation}
and
\begin{equation}
M_{B^*_s}-M_{B_s}\approx M_{B^*}-M_B \; .
\end{equation}
Thus, simply taking
\begin{equation}
\bar \Lambda_s=\bar \Lambda_{u,d}+100 MeV
\end{equation}
seems to be consistent with the heavy quark expansion to $O(\frac {1}{m_Q})$.
 We will adopt this prescription to compare results for mesons containing
 a ``u'' or ``d'' quark with those containing an s quark.

 In summary, our Dirac equation potential is of the form
\begin{equation}
V=\frac {-8\pi}{27r\ln (2.0+\frac {1.87 ({GeV}^{-1})}{r})}+c_\circ
+\gamma^\circ
\sigma r
\end{equation}
where we assumed $\alpha^\infty_s =1$, $N_F=3$, $\Lambda_{\overline
{MS}}=0.240$
; $\sigma$ takes the values 0.25, 0.18 or 0.14 and $c_\circ$ takes values such
 that $\epsilon_{u,d}\equiv \bar \Lambda_{u,d}$ takes a value we select in
 the likely range 150-600 MeV.  For the quark masses we use $m_u=m_d=0$, $
 m_s=0.175 GeV$.  We have checked that this choice for
 $m_s$ is consistent with eq (30).

\section{Quantitative results and discussion}

In this section we use the parametrization for the Dirac equation discussed
in the previous section in order to find the light quark wavefunction, $\psi
_q (\vec x)$.  We then use the formalism developed in section II, to calculate
the Isgur-Wise function $\xi (v.v^\prime )$ in the Breit reference frame, for
different values of the parameters in the potential and for systems where
 the light quark is u,d ($m_u=m_d\approx 0$) or s ($m_s\approx 0.175 GeV$).

 In the case of a central potential, the time-independent Dirac equation for
 a state with angular momentum quantum numbers $j$ and $m$, is reduced to
radial
 equations by writing
\begin{equation}
\psi =\frac {1}{r} \left (\begin{array}{l} g(r)\Omega_{\kappa m}(\theta ,\phi
)\\ -if(r)\Omega_{-\kappa m}(\theta ,\phi )\end{array}\right )
\end{equation}
where $\kappa =-\ell -1$ for $j=\ell +1/2$ and $\kappa =\ell$ for $j=\ell -1/2
$ and $\Omega_{\kappa m}$ is a spinor with spin 1/2 coupled to angular
momentum.  The radial equations are then
\begin{equation}
%\begin{array}{l}
[V_l (r)+m]g(r) +[- {{d}\over {dr}}+ {{\kappa}\over {r}}]f(r)=\epsilon g(r) \;
,
\end{equation}
$$
 [  {{d}\over {dr}}+ {{\kappa}\over {r}}]g(r)+[-m+V_s(r)]f(r)=\epsilon f(r) \;
{}.
%\end{array}
$$
%\end{equation}
In the present calculation the potentials operating on the large and small
components are parametrized as
\begin{equation}
\begin{array}{l}
V_l(r)=\sigma r -\frac {8\pi}{27r\ln [A+B/r]}+c_\circ \; ,  \\
V_s(r)=-\sigma r -\frac {8\pi}{27r\ln [A+B/r]}+c_\circ \; .
\end{array}
\end{equation}

The eigenvalue problem has been solved in two independent ways.
The algebraic approach is to expand $g(r)$ and $f(r)$ in ``basis'' functions
$\phi_i(r)\; ,\; i=1,...,M$ and $\chi_j(r)\; ,\; j=1,...,N$ leading to a
matrix eigenvalue problem of dimension (M+N) for $\epsilon$.
The smallest N eigenvalues
correspond to hole states and the $(N+1)^{\rm st}$ eigenvalue is the lowest
 particle state.  In the present calculation $\phi$ and $\chi$ are taken
 to be
 \begin{equation}
\begin{array}{l}
\phi_i(r)=r^ie^{-\sigma r^2/2}\; ,\; i=1,...,M \; ,\\
\chi_j(r)=r^je^{-\sigma r^2/2}\; ,\; j=1,...,N \; .
\end{array}
\end{equation}
This form is chosen to have the correct analytic behaviour at large r values.
  Matrix elements of the kinetic energy operators and the linear potential
  can be evaluated analytically, but matrix elements of the ``Coulomb'' term
  are computed numerically.  Satisfactory results are obtained for modest
  values of M and N of order 10.

The problem has also been solved fully numerically in the finite difference
approximation by converting the two first order equations to the second
order Pauli equation.  Details of the method used can be found in Ref.
[22].
 Energies found by the two methods are identical.  However, for evaluating
 the form factor the more accurate numerical wavefunctions have been
 employed.

 We present our results for $\xi_{u,d,s}(v.v^\prime )$ in graphical form for
 the range $1<v.v^\prime <4$ which is of phenomenological interest for decays
  of the types $B,B^*\to D,D^*+X\; ; \; B,B^*\to K^* +X\; ; \; D,D^*\to K^* +X
  $.  Figures (1), (2) and (3) exhibit our results for ``inertia'' parameter
  $\bar \Lambda_{u,d}=0.6$ GeV, 0.375 GeV  and 0.15 GeV
   ($\bar \Lambda_s =0.7$ GeV, 0.475 GeV and 0.25 GeV )
  respectively.  For each value of $\bar \Lambda_{u,d}$ we show the result
  $\xi_{u,d}(v.v^\prime )$ for $\sigma =0.25 {GeV}^2 \; , \; 0.18 {GeV}^2 $
   and $0.14 {GeV}^2$ while for each $\bar \Lambda_s$ we give $\xi_s (
   v.v^\prime )$ with $\sigma =0.18 {GeV}^2$ for comparison with $\xi_{u,d}
   (v.v^\prime )$.

In Table 1 we present the values of the zero recoil slope , $\xi^\prime
(v.v^\prime )|_{v.v^\prime =1}$ for the same values of the parameters $\bar
\Lambda_{u,d}\; , \bar \Lambda_s$ and $\sigma$ as used in figures (1-3).

We observe that for $\bar \Lambda_{u,d}\approx 0.15$, $\xi_{u,d} (v.v^\prime
)$ is almost independent of the parameter $\sigma$.  It is clear from eqs
(16-18) that if $\vert \bar\Lambda\vert (=\vert\epsilon_q\vert )$ is very
small,
 $\xi (v.v^\prime )$ is controlled mostly by purely kinematic factors, the
 shape of the wavefunction becomes unimportant (provided that it is properly
 normalized).  We observe also that $\xi_{u,d}(v.v^\prime )$ and $\xi_s(v.v
 ^\prime )$ are quite close ($\sigma =0.18\; , \bar\Lambda_s \approx
 \bar \Lambda_{u,d}+100 MeV$) for all values of $\bar \Lambda_{u,d}$.
 This confirms that the strange quark can be treated as a light quark and
  that $SU(3)_F$ is only softly broken for the processes considered here, once
the
 (phenomenologically imposed) shift in the value of $\bar\Lambda$ has been
 considered.

 Although it is not obvious from figures (1-3), we checked (by using different
 values of $\sigma$ and $\bar \Lambda_{u,d}$ and keeping ${\bar \Lambda_{u,d}
 }^2/\sigma$ constant) that $\xi_{u,d} (v.v^\prime )$ in the range
$1<v.v^\prime
  <4$ depends on the ratio ${\bar \Lambda_{u,d}}^2/\sigma$ only, to a precision
   better than 2\%.  This ``scaling'' is not completely unexpected since
$\sigma
   $ and $\bar \Lambda_{u,d}$ are the main dimensionful parameters entering the
   problem, and $\xi (v.v^\prime )$ is dimensionless.  The other dimensionful
   parameter is $\Lambda_{\overline {MS}}$ in $V_c$, which we are keeping
constant,
  affects $V_c$ only logarithmically and, for massless quarks,  does not much
affect the shape of the
  wavefunction far from the origin (Recall that a pure
  vector-like potential does not produce bound states for massless quarks, and
  we are using $m_u\approx m_d\approx 0$).

  Figure 4 illustrates this scaling effect by giving the values of the zero
  recoil slope $\rho^2\equiv -\xi^\prime (1)$ for different values of
  $\frac {{\bar \Lambda_{u,d}}^2}{\sigma}$.
  The points were obtained from Table 1
  using the 3 values of $\sigma$ and $\bar \Lambda_{u,d}$ quoted.  It
  turns out that all the points satisfy to a very good approximation the
  empirical linear relation
  \begin{equation}
  \rho^2\equiv -\xi^\prime_{u,d}(1)\approx \frac {1}{2}+0.39
  \frac {{\bar \Lambda_{u,d}}^2}{\sigma}
  \end{equation}
  where the numerical coefficient of
  $\frac {{\bar \Lambda_{u,d}}^2}{\sigma}$varies only by $\pm 0.01$ within
  our range of $\bar\Lambda_{u,d}$ and $\sigma$.  This equation can then
  be used to find the zero-recoil slope within our model for an arbitrary
  value of the imput parameters $\bar \Lambda_{u,d}$ and $\sigma$.

  We checked whether this approximate dependence of $\xi (v.v^\prime )$ on
  the ratio
  ${\bar \Lambda_{u,d}}^2/\sigma$ only also holds for the system with a strange
   (light) quark ($m_s\approx 0.175 GeV$).  We found that $\xi_s (v.v^\prime
   )$ does depend on $\sigma$ and $\bar\Lambda_s$ independently, i.e. not
   only through the ratio ${\bar \Lambda_s}^2/\sigma$.  What causes the
   different behaviour in this case is that there is an additional
   dimensionful parameter in the problem, $m_s$, so that we can form two
   independent dimensionless ratios $\bar\Lambda /m_s$ and $\sigma/m_s^2$.
   Regarding $\rho^2_s=-\xi '_s(1)$, we notice from Table 1 that
   consistently $\rho_s^2 > \rho_{u,d}^2$, i.e. the Isgur-Wise function
    has (exhibits) a more rapid decrease for $q\bar Q$ mesons where the
    light quark is an s quark.  This result was obtained as well in the
    calculations of Refs \cite {6,HL}.

   We also studied the effects of changing the value of
   $\Lambda_{\overline {MS}}$
   (B parameter) in $V_c$ (see eq 20, 21 and discussion below) on $\xi (
   v.v^\prime )$, keeping $\sigma$ and $\bar\Lambda$ fixed.  We found that the
   behaviour of $\xi (v.v^\prime )$ is not very sensitive to changes in
$\Lambda
   _{\overline {MS}}$. For instance, using $\sigma =0.18 {GeV}^2$ and $\bar
   \Lambda_{u,d}=0.375 GeV$ ($\bar \Lambda_s=0.475 GeV$), a change of
   $\Lambda_{\overline {MS}}$ by 50\% either way from our central value
   $\Lambda_{\overline {MS}}\approx 0.24 GeV$ changes the zero recoil slope
$\xi^\prime
    (1)$ by only about 2.2\% (3\%).  The relative changes in $(1-\xi (v.
    v^\prime ))$ are of the same order of magnitude for the whole range
    $1<v.v^\prime <4$.

    Finally, we compare our results for $\xi_{u,d} (v.v^\prime )$ with the
    existing experimental data and other theoretical estimates.

As mentioned above, we observed that $\xi_{u,d} (v.v^\prime )$ in our model
depends (to a very good approximation) only on the ratio
${\bar \Lambda_{u,d}}^2/\sigma$.  Hence, we adopt a fitting procedure to the
ARGUS $^,93$\cite {8} and CLEO $^,93$\cite {9} and CLEO '94 \cite {CLEO94}
data in which the parameters of the potential, A=2, B=1.87 $GeV^{-1}$ and
$\sigma =0.18 GeV^2$ are kept fixed, while $\bar
\Lambda_{u,d}$ and the physical observable
$\vert V_{cb}\vert\mu$ are varied in such a way
that $\chi^2$, defined below as being proportional to the total square
deviation from the data (weighed by the experimental uncertainty for each
point) is minimized:
\begin{equation}
\chi^2\equiv \frac {1}{N-2} \sum_{i=1}^N {\left (
\vert V_{cb}\vert\mu \xi [{(v.v^\prime )}_i ]-f_i
\right )}
^2/\sigma_i^2 \; ,
\end{equation}
where
\begin{equation}
\mu \equiv
\sqrt {\frac{\tau_B}{1.48ps}}
\;\;\; \;\;\;({\rm fits\; to \; ARGUS\; '93\; and\; CLEO\; '93})
\end{equation}
and
\begin{equation}
\mu\equiv F(1)\approx \eta_A \;\;\; {\rm in\; the\; notation\; of\;
[11]\; \;\;\;(fits\; to\; CLEO\; '94)}
\end{equation}
The normalization factor $1/(N-2)$ ($N$ experimental points, 2
fitting parameters) allows us to interpret the magnitude of $\chi^2$
for the best fit, $\chi_0^2$, as an absolute measure of the
``quality``of the fit to the corresponding data set.  The factor
$F(1)\approx \eta_A$ required to extract an estimate of $\vert
V_{cb}\vert$ from the CLEO '94 data \cite {CLEO94} has been estimated
theoretically by several authors:
\begin{equation}
\begin{array}{l}
F(1)=0.97\pm 0.04 \;\;\; \cite {1,NUP} \\
F(1)=0.96\pm 0.03 \;\;\; \cite {TM} \\
F(1)<0.94\; ,\; F(1)\approx 0.89\pm 0.03 \;\;\; \cite {SUV}
\; .
\end{array}
\end{equation}

For each data set, we obtain ``acceptable`` ranges for our two
fitting parameters, $\bar\Lambda_{u,d}$ and $\vert V_{cb}\vert \mu$
by requiring
\begin{equation}
\chi^2 < \frac {N}{N-2} \chi_0^2 \; ,
\end{equation}
where $\chi_0^2$ corresponds to the best fit (i.e. $\chi_0^2$ is the
minimum of the function \\ $\chi^2 (\bar\Lambda_{u,d}\; ,\; \vert
V_{cb}\vert\mu ))$.
Because of the scaling
behaviour described above eq (36), the ranges that we obtain for the
parameter $\bar\Lambda_{u,d}$ and fixed $\sigma =0.18\; GeV^2$ can be
translated into ranges for the dimensionless ratio
${\bar\Lambda_{u,d}}^2/\sigma$ where $\sigma$ is allowed to take values
other than $0.18 \; GeV^2$.

We would like to add a cautionary note before giving the results of
our fits.  Because the Isgur-Wise function that we calculate in our
model and use for these fits does {\it not} take into account finite-$M_Q$
corrections ($O(\Lambda_{QCD}^2/M_Q^2)$ at $v.v' =1$ and
$O(\Lambda_{QCD}/M_Q)$ elsewhere), we have to be aware that direct
comparison of our model with experiment (via our $\chi^2$ function)
can be reliable to {\it leading} order in $\Lambda_{QCD}/M_Q$ only.  At
$v.v' =1$, this uncertainty is of order $\Lambda_{QCD}^2/M_Q^2$ and
can be absorbed into the parameter $\mu$ which multiplies $\vert
V_{cb}\vert$ (see eqs 38-40).  However, away from $v.v' =1$ there is
an ``intrinsic`` uncertainty in the comparison of our model to
experiment of expected relative magnitude of order $(v.v'
-1)\Lambda_{QCD}/M_Q$.

The best fit to the ARGUS '93 data (8 points) \cite {8} gives
$\chi^2_0 =0.54$.  Using the above prescription (eq (41)) we obtain
the following acceptable ranges for the fitting parameters:
 $\frac {{\bar\Lambda_{u,d}}^2}{\sigma} =4.8\pm 1.7$ (corresponding to
 (see eq (36)) $\rho_{u,d}^2 =-\xi^\prime_{u,d} (1) =2.4\pm 0.7$) and
 $\vert V_{cb}\vert \sqrt {\frac {\tau_B}{1.48 ps}}=0.050\pm 0.008$.
We note that even if a small value for the string tension $\sigma
=0.14 GeV^2$ were used, the acceptable range for the inertia
parameter would be $\bar\Lambda_{u,d}\approx 0.81\pm 0.15\; GeV$,
which is significantly above most theoretical estimates (see eq
(26)).  The corresponding range for $\rho^2_{u,d}$ overlaps with some
of the theoretical estimates (see Table 2) but is centered above most
of the predicted ranges.

The best fit to the CLEO '93 data (7 points) \cite {9} is poorer with
$\chi^2_0\approx 1.18$.  Here we find the ranges
 $\frac {{\bar\Lambda_{u,d}}^2}{\sigma} =3.4\pm 1.8$ (corresponding to
 (see eq (36)) $\rho_{u,d}^2 =-\xi^\prime_{u,d} (1) =1.8\pm 0.7$) and
 $\vert V_{cb}\vert \sqrt {\frac {\tau_B}{1.48 ps}}=0.043\pm 0.008$.
There is some overlap between the resulting range for
$\bar\Lambda_{u,d}$ (eg $\bar\Lambda_{u,d}=0.66\pm 0.19$ for $\sigma
=0.14 \; GeV^2$) and independent theoretical estimates (See eq 26).
Also, the range for the slope $\rho^2_{u,d}$ significantly overlaps
with several previous theoretical predictions (See Table 2).  We
would like to remark that if we ignore the CLEO '93 data point
corresponding to highest recoil ($v.v' \approx 1.5$), the fit to the
remaining 6 points is greatly improved.  We obtain in this case
$\chi^2_0=0.62$ and ranges
 $\frac {{\bar\Lambda_{u,d}}^2}{\sigma} =2.1\pm 1.6$ (corresponding to
 $\rho_{u,d}^2 =1.3\pm 0.6$) and
 $\vert V_{cb}\vert \sqrt {\frac {\tau_B}{1.48 ps}}=0.038\pm 0.006$.
For commonly used values for the string tension ($\sigma =0.14-0.18
\; GeV^2$) the acceptable range for $\bar\Lambda_{u,d}$ in this case
is centered well within the range of previous theoretical estimates
(See eq 26).  The range for the zero recoil slope
$\rho^2_{u,d}=-\xi_{u,d}^\prime (1)$ is very similar to the ranges
obtained in recent lattice estimates \cite {24,25} as well as from
several other theoretical calculations (See Table 2).

The best fit of our model to the recent CLEO '94 data analysis (7
points) \cite {CLEO94} gives $\chi^2_0 =0.50$ which is the lowest
$\chi_0^2$ of all our fits, in spite of the smaller experimental
error bars.  This means that within a region of its parameter space,
our model agrees well with this data set.  In turn, we expect that
this data with smaller error bars will be useful in selecting a
relatively narrow region for our model parameters (mainly
$\bar\Lambda_{u,d}^2/\sigma$) as well as for the Standard Model
flavour mixing parameter $\vert V_{cb}\vert$ (or at least the product
$\vert V_{cb}\vert F(1)$).  The parameter ranges obtained by imposing
eq (41) are:
 $\frac {{\bar\Lambda_{u,d}}^2}{\sigma} =1.9\pm 0.7$ (corresponding to
 $\rho_{u,d}^2 =-\xi'_{u,d}(1)=1.3\pm 0.3$) and
 $\vert V_{cb}\vert F(1)=0.037\pm 0.003$. As expected, the smaller
experimental error bars lead to a better determination of our model
parameter $\bar\Lambda_{u,d}^2/\sigma$ (and correspondingly, via eq
(36), of the slope at zero recoil $\rho^2_{u,d}$) as well as of the
physically interesting product $\vert V_{cb}\vert F(1)$.
We note that for the commonly used values for the string tension
$\sigma = 0.14-0.18\; GeV^2$, the resulting range for
$\bar\Lambda_{u,d}$ ($0.41 \; GeV <\bar\Lambda_{u,d}<0.69\; GeV$)
overlaps significantly with the upper half of the range given in eq
(26), which was obtained from recent theoretical estimates of $m_b$
\cite {3,4,5}.  In this respect, the larger values of $\sigma$ (eg
$\sigma =0.25\; GeV^2$ (See discussion below eq (22)), seem to be
less favoured.  We would like to remark here that an estimate of the
pseudoscalar decay constants $f_D$ and $f_B$ in the context of the
relativistic model used here \cite {WIP}, when compared with recent
estimates of these constants with lattice and QCD Sum Rule methods,
favours the lower values for $\sigma$ as well.

Because the Isgur-Wise function is normalized at zero recoil, $\xi
(1)\equiv 1$, the slope at zero recoil $\xi' (1)$ determines to a
good approximation the value of the function close to $v.v' =1$.
Therefore, at least close to $v.v' =1$ (all the existing data is in
the interval [1,1.5]) the slope $\xi' (1) \equiv -\rho^2$ is a
reliable tool for comparison of the different theoretical estimates of $\xi
(v.v' )$.  Our original model parameter ranges $0.15 \; GeV
<\bar\Lambda_{u,d}<0.6\; GeV$ and $0.14\; GeV^2<\sigma <0.25 \;
GeV^2$ (See discussion between eqs (22) and (26)) give (through our
result in eq (36)) a wide range $0.54 <\rho^2_{u,d}<1.5$, which
overlaps with many different theoretical estimates (See Table 2 and
refs [6-8] and [27-38]).  The only exceptions are refs [33-36].  On
the other hand, the relatively narrow range for $\rho_{u,d}^2$
favoured by our fit to the CLEO '94 data \cite {CLEO94}
($\rho_{u,d}^2 =1.3\pm 0.3$), overlaps with the predictions in refs
[7], [8], and [28-32] only, although is not far from the range
advocated in ref [6] (See Table 2).  Both the lattice gauge theory
calculations [31] and [32] contain the range favoured by the fit of
our model to the CLEO '94 data within their predicted ranges.

The extraction of a precise value for the Standard Model parameter
$\vert V_{cb}\vert$ from the range $\vert V_{cb}\vert F(1)=0.037\pm
0.003$ favoured by our fit to the CLEO '94 data is partially hindered
by the present uncertainty in the theoretical determination of
$F(1)$.  For example, if we use the whole range of values for $F(1)$
given in eq (40), we would obtain $0.034 <\vert V_{cb}\vert <0.046$.

It is interesting to note that our fits to the CLEO '94 data (7
points) produce similar central values for
$\bar\Lambda_{u,d}^2/\sigma$, $\rho^2_{u,d}$ and $\vert V_{cb}\vert$
(for $F(1)\leq 1$) as do our fits to the CLEO '93 data with the point
corresponding to the largest $v.v'$ omitted from the CLEO '93 set
(i.e. 6 points).  The resulting ranges for these quantities are,
however, significantly narrower (by a factor of about 2) for the CLEO
'94 data set.  We should also point out that the above mentioned
central values for $\rho^2_{u,d}$ and $\vert V_{cb}\vert F(1)$ that
we obtained from the fits of our model to the CLEO '94 data set (Fig
12 in ref \cite {CLEO94}) are both somewhat larger than the central values for
these quantities obtained in the CLEO analysis carried out in ref
\cite {CLEO94} (In their notation $a^2=\rho^2_{u,d}$).  Once the
uncertainties are included, however, our results for these quantities
are compatible.

Eventually, when the experimental data becomes more precise and the
$1/M_Q$ corrections can be incorporated with fewer uncertainties, one
should be able to further narrow the allowed range for our main model
input, $\bar\Lambda_{u,d}/\sigma$, as well as for the flavour mixing
parameter $\vert V_{cb}\vert$.

\section{Acknowledgements}
We would like to thank J. Sloan, C. Quigg, I. Bigi and B. Blok for useful
discussions.  This work was supported in part by the Natural Sciences
and Engineering Research Council of Canada.

\newpage
\begin{flushleft}
\Large \bf
Table Captions
\end{flushleft}
{\Large \bf Table 1:}
The slope of the Isgur-Wise function at zero recoil for various
values of the parameters $\bar \Lambda$ and $\sigma$. \\
\vskip 1cm
\hskip -0.6cm
{\Large \bf Table 2:}
Various theoretical estimates of $\rho^2_{u,d} =-\xi^\prime (1)$.

\newpage
\begin{flushleft}
\Large \bf Figure Captions
\end{flushleft}
{\Large \bf Figure 1:}
The Isgur-Wise function $\xi_{u,d}$ ($\xi_s$) for $\bar\Lambda_{u,d}
=0.15 \; GeV$ ($\bar\Lambda_s =0.25 \; GeV$) and $\sigma =0.25\; , 0.18\;
{\rm and}
\; 0.14 \; {GeV}^2$ ($\sigma =0.18 \; {GeV}^2$). \\
\vskip 1cm
\hskip -0.6cm
{\Large \bf Figure 2:}
The Isgur-Wise function $\xi_{u,d}$ ($\xi_s$) for $\bar\Lambda_{u,d}
=0.375 \; GeV$ ($\bar\Lambda_s =0.475 \; GeV$) and $\sigma =0.25\; , 0.18\;
{\rm and}
\; 0.14 \; {GeV}^2$ ($\sigma =0.18 \; {GeV}^2$). \\
\vskip 1cm
\hskip -0.6cm
{\Large \bf Figure 3:}
The Isgur-Wise function $\xi_{u,d}$ ($\xi_s$) for $\bar\Lambda_{u,d}
=0.6 \; GeV$ ($\bar\Lambda_s =0.7 \; GeV$) and $\sigma =0.25\; , 0.18\;
{\rm and}
\; 0.14 \; {GeV}^2$ ($\sigma =0.18 \; {GeV}^2$). \\
\vskip 1cm
\hskip -0.6cm
{\Large \bf Figure 4:}
The slope of the Isgur-Wise function at zero recoil
$\rho^2_{u,d}=-\xi^\prime_{u,d} (1)$ for different values of the
parameter $\frac {{\bar\Lambda_{u,d}}^2}{\sigma}$.

\newpage

\vskip 2cm
\hskip -2.5cm
{\hskip 6cm \bf \Large Table 1} \\
{
\begin{tabular}{c|c|c|c}
\hline
\multicolumn{1}{|c|}{             } &
\multicolumn{1}{c|}{$\sigma =0.25$} &
\multicolumn{1}{c|}{$\sigma =0.18$} &
\multicolumn{1}{c|}{$\sigma =0.14$} \cr
\hline
\multicolumn{1}{|c|}{
$\begin{array}{l} \bar\Lambda_{u,d}=0.15 \\
\bar\Lambda_s=0.25 \end{array}$} &
\multicolumn{1}{c|}{
$\begin{array}{l} \rho^2_{u,d}= 0.53 \\
\rho^2_s=0.58 \end{array}$} &
\multicolumn{1}{c|}{
$\begin{array}{l} \rho^2_{u,d}=0.55 \\
\rho^2_s= 0.61 \end{array}$} &
\multicolumn{1}{c|}{
$\begin{array}{l} \rho^2_{u,d}=0.56 \\
\rho^2_s=0.64 \end{array}$} \cr
\hline
\multicolumn{1}{|c|}{
$\begin{array}{l} \bar\Lambda_{u,d}= 0.375 \\
\bar\Lambda_s =0.475 \end{array}$} &
\multicolumn{1}{c|}{
$\begin{array}{l} \rho^2_{u,d}= 0.72 \\
\rho^2_s=0.80 \end{array}$} &
\multicolumn{1}{c|}{
$\begin{array}{l} \rho^2_{u,d}=0.81  \\
\rho^2_s=0.90 \end{array}$} &
\multicolumn{1}{c|}{
$\begin{array}{l} \rho^2_{u,d}= 0.89 \\
\rho^2_s=0.99 \end{array}$} \cr
\hline
\multicolumn{1}{|c|}{
$\begin{array}{l} \bar\Lambda_{u,d}=0.6 \\
\bar\Lambda_s =0.7 \end{array}$} &
\multicolumn{1}{c|}{
$\begin{array}{l} \rho^2_{u,d}= 1.06 \\
\rho^2_s=1.16 \end{array}$} &
\multicolumn{1}{c|}{
$\begin{array}{l} \rho^2_{u,d}= 1.28 \\
\rho^2_s= 1.37 \end{array}$} &
\multicolumn{1}{c|}{
$\begin{array}{l} \rho^2_{u,d}= 1.50 \\
\rho^2_s=1.56 \end{array}$} \cr
\hline
\end{tabular}
\hskip 2.5cm

\vskip 2cm
\hskip -2.5cm
{\hskip 5cm \bf \Large Table 2} \\
{
\begin{tabular}{c|c}
\hline
\multicolumn{1}{|c|}{Bjorken [8]} &
\multicolumn{1}{c|}{ $>$ $  {1} \over 4$} \cr
\hline
\multicolumn{1}{|c|}{Isgur et al. [27]} &
\multicolumn{1}{c|}{$ 0.63 (0.33)$} \cr
\hline
\multicolumn{1}{|c|}{Rosner [28]} &
\multicolumn{1}{c|}{$1.44\pm 0.41$} \cr
\hline
\multicolumn{1}{|c|}{Mannel et al. [29]} &
\multicolumn{1}{c|}{$1.77\pm 0.74$} \cr
\hline
\multicolumn{1}{|c|}{Neubert [30]} &
\multicolumn{1}{c|}{$1.28\pm 0.25$} \cr
\hline
\multicolumn{1}{|c|}{Bernard et al. [31]} &
\multicolumn{1}{c|}{$1.41\pm 0.19\pm 0.41$} \cr
\hline
\multicolumn{1}{|c|}{UKQCD Collaboration [32]} &
\multicolumn{1}{c|}{$1.2 ^{+7}_{-8}$} \cr
\hline
\multicolumn{1}{|c|}{Radyushkin [33]} &
\multicolumn{1}{c|}{$\infty$} \cr
\hline
\multicolumn{1}{|c|}{Karanikas and Ktorides [34]} &
\multicolumn{1}{c|}{$0$} \cr
\hline
\multicolumn{1}{|c|}{Sadzikowski and Zalewski [7]} &
\multicolumn{1}{c|}{$1.24$} \cr
\hline
\multicolumn{1}{|c|}{Kugo et al. [35]} &
\multicolumn{1}{c|}{$1.8-2.0$} \cr
\hline
\multicolumn{1}{|c|}{Ivanov et al. [36]} &
\multicolumn{1}{c|}{$0.43$} \cr
\hline
\multicolumn{1}{|c|}{Ivanov and Mizutani [37]} &
\multicolumn{1}{c|}{$0.42-0.82$} \cr
\hline
\multicolumn{1}{|c|}{Narison [6]} &
\multicolumn{1}{c|}{$0.52-0.92$} \cr
\hline
\multicolumn{1}{|c|}{Blok and Shifman [38]} &
\multicolumn{1}{c|}{$0.5-0.8$} \cr
\hline
\end{tabular}
\hskip 2.5cm


\begin{thebibliography}{99}
\bibitem{10} We remark that in Ref (7), a similar expression was obtained,
except for a factor of $\gamma^{-1}$ in their complex phase.  That factor
can be traced to an error in the Lorentz transformation in their eq (21).
\bibitem{NUP} Ref. [1] gives $F=0.986\pm 0.04$ ($\hat\xi (1)=1.00\pm
0.04$) (Neubert writes the form factor $F(y)$ as $\eta_A\hat\xi (y)$,
where $\eta_A$ is a calculable radiative correction for hard gluon
processes ($\eta_A =0.986\pm 0.006$ [1]) and $\hat\xi (y)$is the
non-perturbative part of the form factor), but the final version, to
appear in Physics Reports, will have $\hat\xi (1)=0.98\pm 0.04$,
leading to $F(1)=0.97\pm 0.04$.
\end{thebibliography}
\end{document}